\documentstyle[12pt]{article}
\begin{document}
\begin{titlepage}
\title{On the exponentially large probability of transition through the
Lavrelashvili-Rubakov-Tinyakov wormhole}

\author{
   O.Yu. Shvedov  \\
{\small{\em Institute for Nuclear Research of the Russian
Academy of Sciences,  }}\\ {\small{\em 60-th October Anniversary Prospect
7a, Moscow 117312, Russia
}}\\ {\small and}\\ 
{\small {\em Sub-faculty of Quantum Statistics and Field Theory,}}\\
{\small{\em Department of Physics, Moscow State University }}\\
{\small{\em Vorobievy gory, Moscow 119899, Russia}} }

\end{titlepage}
\maketitle

\begin{center}
{\bf Abstract}
\end{center}

The model consisting of gravitational, scalar and axionic fields 
is considered.
It is shown that the action of the 
Lavrelashvili-Rubakov-Tinyakov wormhole can be made 
arbitrarily negative by varying the parameters of the model. This means that
semiclassically calculated 
probability of transition through this wormhole is not exponentially
small (as usual) but exponentially large.
\begin{center}
{\bf PREPRINT gr-qc/9602049}
\end{center}
\newpage

Wormholes -- solutions to the euclidean Einstein equations that  connect
two asymptotically flat regions -- are  very important for semiclassical
calculations of probabilities of topology change transitions in quantum 
gravity. An example of wormhole is the Giddings-Strominger euclidean solution
\cite{GS} in the model of the axionic field minimally coupled to gravity. The
action $\cal S$ of this wormhole is positive and large at large values of the
axionic charge flowing through the wormhole. The probability of the topology
change transition which is proportional to
$\exp(-{\cal S})$ is then small, so
that the wormhole transition is suppressed.

It is shown in this paper that there exist such models that the action of the
wormhole is not positive but negative. Moreover, by varying the parameters of 
the model one can make the absolute value of the action as large as one wants.
This implies that formally calculated semiclassical 
probability $e^{-{\cal S}}=e^{+|{\cal S}|}$ of the wormhole
transition
will be exponentially large, so that the dilute-wormhole-gas approximation
used usually for calculation \cite{C,GS2} of the wormhole effects will be
not applicable.

Namely, consider the 
Lavrelashvili-Rubakov-Tinyakov model \cite{LRT,RT1,RT2} consisting of
the scalar field $\varphi$ and axionic field $H_{\mu\nu\lambda}=
\partial_{\mu}B_{\nu\lambda}+\partial_{\nu}B_{\lambda\mu}+
\partial_{\lambda}B_{\mu\nu}$ ($B$ is an antisymmetric tensor) which are 
minimally coupled to gravity. The euclidean action of the model is
\begin{equation}
\label{1}
{\cal S}=\int d^4x \sqrt{g} \left( -\frac{1}{2\kappa} R + \frac{1}{2}
g^{\mu\nu}\partial_{\mu}\varphi\partial_{\nu}\varphi +\frac{\mu^2}{\kappa}
V(\sqrt{\kappa}\varphi) + \frac{1}{12} H_{\mu\nu\lambda}H^{\mu\nu\lambda}
\right),
\end{equation}
where $\kappa$ is a gravitational coupling constant of the dimensions 
$Gev^{-2}$, $\mu$ has the dimensions $Gev$. The dependence of the scalar 
potential on the parameters of the model is chosen in this way for the 
following aim. If one calculates the probability of the euclidean transition, 
one should consider \cite{H} the path integral of $\exp(-{\cal S})$. By
the rescaling
$$
\Phi=\varphi\sqrt{\kappa}, \tilde{x}=\mu x, \tilde{H}_{\alpha\beta\gamma}=
H_{\alpha\beta\gamma}\sqrt{\kappa/\mu^2}, S={\cal S} \kappa\mu^2
$$
one brings this integral to the saddle-point form
\begin{equation}
\int DgD\Phi D\tilde{H} \exp\left(-\frac{1}{\mu^2\kappa}S\right)
\label{2}
\end{equation}
if $\mu^2\kappa \ll 1$. Note that the dependence like 
$V(\sqrt{\kappa}\varphi)/\kappa$ is often considered in investigations of 
large-order behaviour of perturbation theory (for a review see \cite{ZJ}), as
well as in the Maslov complex-WKB theory \cite{M}.

The saddle point of the integral (\ref{2}) which is the solution to the
classical field equations has the following 
spherically-symmetric form \cite{LRT,RT1,RT2}
$$
\Phi=\Phi(\tilde{x}^0), \tilde{H}_{0ij}=0, 
\tilde{H}_{123}=\frac{q}{2\pi^2}\sqrt{det \gamma},
$$
$$
\tilde{ds}^2=g_{\mu\nu} d\tilde{x}^{\mu} d\tilde{x}^{\nu} = 
(d\tilde{x}^0)^2 + a^2(\tilde{x}^0) d\Omega_3^2
$$
(where $d\Omega_3^2=\gamma_{11}(d\tilde{x}^1)^2
+\gamma_{22}(d\tilde{x}^2)^2+\gamma_{33}(d\tilde{x}^3)^2$ 
is the metric of the unit 3-sphere)
if the following equations for $\Phi(\tau)$ and $a(\tau)$ are satisfied:
\begin{equation}
(a^3\dot{\Phi})\dot{} = a^3 V'(\Phi),
\label{3a}
\end{equation}
\begin{equation}
\dot{a}^2=1+ \frac{1}{3}a^2 (\frac{1}{2} \dot{\Phi}^2 -V(\Phi)) -
\frac{q^2}{24\pi^4a^4}.
\label{3b}
\end{equation}
It is also required that the potential $V(\Phi)$ has a local minimum at
some point $\Phi_0$, where $V(\Phi_0)=0$,
and the following boundary conditions are imposed
on the semiwormhole:
\begin{equation}
\dot{a}(0)=\dot{\Phi}(0)=0, \Phi(+\infty)=\Phi_0
\label{4}
\end{equation}
(this implies that the solution can be continued into a region $\tau<0$
in such a way that $a(-\tau)=a(\tau), \Phi(-\tau)=\Phi(\tau))$.
The quantity $S/2$ being proportianal to the action 
${\cal S}/2$ of the semiwormhole
is
\begin{equation}
S/2 = \int_0^{\infty} d\tau [-6\pi^2a(1-a\ddot{a}-\dot{a}^2)+
2\pi^2a^3(\frac{1}{2} \dot{\Phi}^2 +V(\Phi))+\frac{q^2}{4\pi^2a^3} ].
\label{5}
\end{equation}
Note also that the quantity $q/(\sqrt{\kappa}\mu^2)$ is the axionic
charge flowing through the wormhole.

Let us show that for some potentials $V$ there exist wormholes with
negative action. To prove this statement, it is convenient to reformulate
the problem. Let us be interested not
in construction of the solution to eqs.
(\ref{3a}), (\ref{3b}) that obeys eq.(\ref{4}) but in finding the functions
$\Phi(\tau)$ and $V(\Phi)$ if the function $a(\tau)$ is given. It will be
then not hard to choose the latter function in such a way that the quantity
(\ref{5}) will be negative.

By differentiating eq.(\ref{3b}) with respect to $\tau$ and making use of
eq.(\ref{3a}), one obtains that
\begin{equation}
\dot{\Phi}^2 = -2\frac{d}{d\tau}
\left(\frac{\dot{a}}{a}\right) - \frac{2}{a^2}
+\frac{q^2}{4\pi^4a^6}.
\label{6a}
\end{equation}
It follows then from eq.(\ref{3b}) that
\begin{equation}
V(\Phi) = -\frac{d}{d\tau}
\left(\frac{\dot{a}}{a}\right) + \frac{2}{a^2}
-3\left(\frac{\dot{a}}{a}\right)^2.
\label{6b}
\end{equation}
Finally, the action (\ref{5}) takes the following form:
\begin{equation}
S=\int_0^{\infty} d\tau \left[
\frac{q^2}{2\pi^2a^3}- 4\pi^2a +2\pi^2(a^2\dot{a})\dot{}\right]
\label{7}
\end{equation}
Oppositely, eqs. (\ref{6a}), (\ref{6b}) also imply eqs.(\ref{3a}),
(\ref{3b}) if $\dot{\Phi}\ne 0$.

Let us consider the function $a$ shown in fig.1. When $\tau\gg\tau_0$,this
function can be approximated by the Giddings-Strominger wormhole solution
\cite{GS} defined from the relation
$$
\dot{a}_{GS}^2=1-\frac{q^2}{24\pi^4a_{GS}^4}
$$
and condition
$$
\dot{a}_{GS}(\tau_0)=0.
$$
Nevertheless, one should be careful in constructing the function $a$, since
the dependence $$V(\Phi)\sim \alpha^2(\Phi-\Phi_0)^2/2$$ (as
$\Phi \rightarrow\Phi_0$) of the potential $V$ should be taken into
account (the parameter $\alpha$ is proportional to the mass of the 
scalar particle $\mu\alpha$).
However, let us choose the function $a$ as
\begin{equation}
a(\tau) \sim a_{GS}(\tau)+\epsilon b(\tau) +O(\epsilon^2), \tau\gg \tau_0,
\label{8}
\end{equation}
where $\epsilon$ is a small parameter, while the function $b(\tau)$
is found from the equation
$$
2\dot{a}_{GS}\dot{b} =\frac{1}{3} a_{GS}^2 \left(\frac{1}{2}
\dot{\phi}_{lin}^2 - \frac{\alpha^2}{2} \phi_{lin}^2 \right)
+\frac{q^4}{4\pi^4a_{GS}^5}b ; b(\infty)=0,
$$
where $\phi_{lin}$ is a solution to the linearized equation (\ref{3a})
$$
(a_{GS}^3\dot{\phi}_{lin})\dot{}=a_{GS}^3 \alpha^2 \phi_{lin}.
$$
One can show that the large-$\tau$-behaviour (\ref{8}) leads to the correct
mass term of the potential (\ref{6b}).

Consider now the region $\tau_1< \tau< \tau_0+\delta$, where $\delta$ is a
finite quantity, while $\tau_1$ is found from the relation
$$
a_{GS}(\tau_1)=3a_1=3(q^2/8\pi^4)^{1/4}, \dot{a}_{GS}(\tau_1) <0.
$$
Let us choose the function $a(\tau)$ in this region to be also slightly
different from $a_{GS}(\tau)$,
$$
|a(\tau)-a_{GS}(\tau)| \ll a_{GS}(\tau),
$$
while the quantity (\ref{6a}) is required to be positive, and
\begin{equation}
|\dot{a}(\tau_1)-\dot{a}_{GS}(\tau_1)| \ll \dot{a}_{GS}(\tau_1),
a(\tau_1)> 2a_1
\label{8a}
\end{equation}
(see fig.1). These conditions can be satisfied if the parameter $\epsilon$ in
eq.(\ref{8}) is sufficiently small.

To estimate the contribution of the region $\tau>\tau_1$ to the action
(\ref{8}) , notice that for the Giddings-Strominger wormhole such
quantity is lesser than
$
\pi q\sqrt{6}/2.
$
This means that for sufficiently small ratio
$|a(\tau)-a_{GS}(\tau)| / a_{GS}(\tau)$ the estimation
\begin{equation}
S=\int_{\tau_1}^{\infty} d\tau \left[
\frac{q^2}{2\pi^2a^3}- 4\pi^2a +2\pi^2(a^2\dot{a})\dot{}\right]
\le \frac{\pi q\sqrt{6}}{2}
\label{9}
\end{equation}
will take place.

Consider now the region $0< \tau< \tau_1$. Let us choose the function $a$
as shown in fig.1 in such a way that the conditions:

(i) the quantity (\ref{6a}) is positive;

(ii) $\dot{a}(0)=0, \dot{\Phi}(0)=0$

are satisfied. One can do it, because the first condition means that
\begin{equation}
\frac{d}{d\tau}\left(\left(\frac{\dot{a}}{a}\right)^2
-\frac{1}{a^2} +\frac{q^2}{24\pi^4a^6}
\right) >0
\label{9a}
\end{equation}
(since $a$ decreases) and does not contradict the boundary conditions,
while the condition (ii) is satisfied when
$$
a(\tau)=a(0)-\frac{\tau^2}{2}\left(\frac{1}{a(0)}-\frac{q^2}{8\pi^4 a^5(0)}
\right)-\gamma\tau^4 + o(\tau^4)
$$
at small $\tau$ ($\gamma$ is sufficiently large to provide eq.(\ref{9a})).

The conribution of this region to the action can be made arbitrarily negative.
Namely, it follows from eq.(\ref{8a}) that
\begin{equation}
\int_0^{\tau_1} d\tau \left[
\frac{q^2}{2\pi^2a^3}- 4\pi^2a +2\pi^2(a^2\dot{a})\dot{}\right] <
-\tau_1 \frac{15\pi}{2\cdot 8^{1/4}} \sqrt{q}.
\label{10}
\end{equation}
As the parameter $\tau_1$ can be made arbitrarily large, the absolute value
of the quantity (\ref{10}) will exceed then the finite quantity (\ref{9}).
Therefore, the total action (\ref{7}) will be negative.

When the function $a$ is constructed, one can easily define $\Phi(\tau)$
by using the function $\dot{\Phi}>0$ found from eq.(\ref{6a}) and
making use of the boundary condition $\Phi(\infty)=\Phi_0$. The function
$V(\Phi(\tau)),\tau>0$, as well as $V(\Phi)$, $\Phi(0)<\Phi<\Phi_0$,
is defined from eq.(\ref{6b}). Thus, we see that $S<0$ for some potential
$V(\Phi)$.

Note that one can also investigate in analogous way the case, when the
axionic field $H$ is replaced by the massless scalar field. The wormhole
solution with the purely imaginary value of the massless field also exists,
while the equations of motion with given value of the global charge
\cite{L,CL} of the massless field coincides with eqs.(\ref{3a}),
(\ref{3b}).

Thus, we have seen that the wormhole action can be made negative. This implies
that the contribution of the saddle point under consideration to the
functional integral (\ref{2}) is
$$
\exp(\frac{1}{\mu^2\kappa} |S|).
$$
Notice that the semiclassical approximation is valid if the parameter
$\mu^2\kappa$ entering to eq.(\ref{2}) and playing the role of the Planck
constant is small. For example, if one considers $\mu$ to be 
of order of masses of ordinary particles
 ($\mu\sim 1Gev$) then the quantity $\mu^2\kappa$
will be $10^{-38}$ -- a very good small parameter of the expansion.
However, the semiclassically evaluated
probability of the wormhole transition will be exponentially
large
\begin{equation}
10^{10^{38}}.
\label{11}
\end{equation}
If we adopt the point of view of \cite{C,GS2} that wormholes are relevant
to the values of the coupling constants, we will find that such values are
also of order eq.(\ref{11}).

Therefore, the obtained result imposes a restriction on grand unified models:
one should avoid such potentials that allow wormholes with negative actions.
Another possible resolvation of the difficulty is to introduce the
topological term \cite{GS}, add it to the action and suppose the
topological coupling constant to be large.

Note also that in ordinary particle physics the considered scalar field
$\varphi$ behaves as a massive field of the mass $\alpha\mu$
without self-interaction (since the coefficient of the $\varphi^n$-term
which is proportional to $\mu^2\kappa^{n/2-1} \sim (10^{-19})^{n-2}
Gev^{4-n}$ is small). The dependence of $V$ on $\varphi$ may be
experimentally found only if one works with particles with Planck energies.
We are faced then with an interesting situation: the only influence
of the Planck-scale physics on the particle physics is not the perturbation
theory processes but tunneling processes being undamped.

The author is indebted to G.V.Lavrelashvili, V.A.Rubakov and P.G.Tinyakov  
 for helpful
discussions. This work was supported in part by ISF, grant \# MKT300.

\newpage
{\bf Figure caption.}

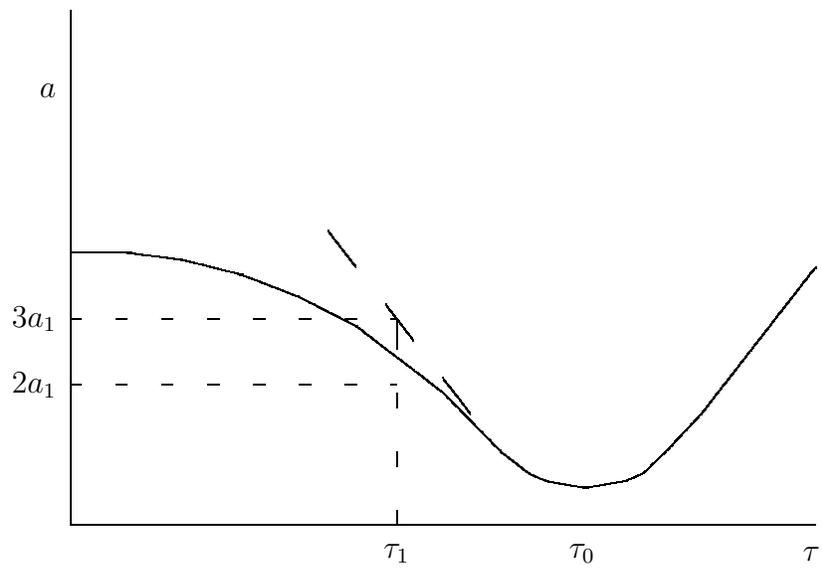
{\bf Fig.1.} The function $a(\tau)$ (solid line) and its approximation
$a_{GS}(\tau)$ (dashed line). The region $\tau>\tau_1$ gives rise to the
finite positive contribution to the wormhole action, eq.(\ref{7}),
while the contribution of the region $0<\tau<\tau_1$ can be made arbitrarily
negative.
\newpage
\setlength{\unitlength}{0.240900pt}
\ifx\plotpoint\undefined\newsavebox{\plotpoint}\fi
\sbox{\plotpoint}{\rule[-0.200pt]{0.400pt}{0.400pt}}%
\begin{picture}(1500,600)(0,0)
\font\gnuplot=cmr10 at 10pt
\gnuplot
\put(176,68){\usebox{\plotpoint}}
\put(176.0,68.0){\rule[-0.200pt]{281.853pt}{0.400pt}}
\put(176,68){\usebox{\plotpoint}}
\put(176.0,68.0){\rule[-0.200pt]{0.400pt}{194.888pt}}
\put(689,68){\usebox{\plotpoint}}
\put(689.0,68.0){\rule[-0.200pt]{0.400pt}{5.541pt}}
\put(689.0,160.0){\rule[-0.200pt]{0.400pt}{5.782pt}}
\put(689.0,253.0){\rule[-0.200pt]{0.400pt}{5.541pt}}
\put(689.0,345.0){\rule[-0.200pt]{0.400pt}{11.322pt}}
\put(176.0,288.0){\rule[-0.200pt]{4.336pt}{0.400pt}}
\put(248.0,288.0){\rule[-0.200pt]{4.336pt}{0.400pt}}
\put(320.0,288.0){\rule[-0.200pt]{4.336pt}{0.400pt}}
\put(392.0,288.0){\rule[-0.200pt]{4.336pt}{0.400pt}}
\put(464.0,288.0){\rule[-0.200pt]{4.336pt}{0.400pt}}
\put(536.0,288.0){\rule[-0.200pt]{4.336pt}{0.400pt}}
\put(608.0,288.0){\rule[-0.200pt]{4.336pt}{0.400pt}}
\put(680,288){\usebox{\plotpoint}}
\put(680.0,288.0){\rule[-0.200pt]{2.168pt}{0.400pt}}
\put(176,392){\usebox{\plotpoint}}
\put(176.0,392.0){\rule[-0.200pt]{4.336pt}{0.400pt}}
\put(248,392){\usebox{\plotpoint}}
\put(248.0,392.0){\rule[-0.200pt]{4.336pt}{0.400pt}}
\put(320,392){\usebox{\plotpoint}}
\put(320.0,392.0){\rule[-0.200pt]{4.336pt}{0.400pt}}
\put(392,392){\usebox{\plotpoint}}
\put(392.0,392.0){\rule[-0.200pt]{4.336pt}{0.400pt}}
\put(464,392){\usebox{\plotpoint}}
\put(464.0,392.0){\rule[-0.200pt]{4.336pt}{0.400pt}}
\put(536,392){\usebox{\plotpoint}}
\put(536.0,392.0){\rule[-0.200pt]{4.336pt}{0.400pt}}
\put(608,392){\usebox{\plotpoint}}
\put(608.0,392.0){\rule[-0.200pt]{4.336pt}{0.400pt}}
\put(680,392){\usebox{\plotpoint}}
\put(680.0,392.0){\rule[-0.200pt]{2.168pt}{0.400pt}}
\put(581,530){\usebox{\plotpoint}}
\multiput(581.58,527.44)(0.498,-0.645){87}{\rule{0.120pt}{0.616pt}}
\put(671,415){\usebox{\plotpoint}}
\multiput(671.58,412.44)(0.498,-0.645){87}{\rule{0.120pt}{0.616pt}}
\put(761,299){\usebox{\plotpoint}}
\multiput(761.58,296.44)(0.498,-0.645){87}{\rule{0.120pt}{0.616pt}}
\put(1346,472){\usebox{\plotpoint}}
\multiput(1344.92,469.45)(-0.500,-0.642){357}{\rule{0.120pt}{0.613pt}}
\multiput(1345.17,470.73)(-180.000,-229.727){2}{\rule{0.400pt}{0.307pt}}
\multiput(1164.92,238.83)(-0.498,-0.527){105}{\rule{0.120pt}{0.522pt}}
\multiput(1165.17,239.92)(-54.000,-55.916){2}{\rule{0.400pt}{0.261pt}}
\multiput(1109.88,182.92)(-0.513,-0.498){67}{\rule{0.511pt}{0.120pt}}
\multiput(1110.94,183.17)(-34.939,-35.000){2}{\rule{0.256pt}{0.400pt}}
\multiput(1071.85,147.92)(-1.142,-0.492){21}{\rule{1.000pt}{0.119pt}}
\multiput(1073.92,148.17)(-24.924,-12.000){2}{\rule{0.500pt}{0.400pt}}
\multiput(1039.08,135.92)(-2.949,-0.492){19}{\rule{2.391pt}{0.118pt}}
\multiput(1044.04,136.17)(-58.038,-11.000){2}{\rule{1.195pt}{0.400pt}}
\multiput(976.08,126.58)(-2.949,0.492){19}{\rule{2.391pt}{0.118pt}}
\multiput(981.04,125.17)(-58.038,11.000){2}{\rule{1.195pt}{0.400pt}}
\multiput(918.85,137.58)(-1.142,0.492){21}{\rule{1.000pt}{0.119pt}}
\multiput(920.92,136.17)(-24.924,12.000){2}{\rule{0.500pt}{0.400pt}}
\multiput(893.45,149.58)(-0.643,0.498){67}{\rule{0.614pt}{0.120pt}}
\multiput(894.73,148.17)(-43.725,35.000){2}{\rule{0.307pt}{0.400pt}}
\multiput(849.92,184.00)(-0.498,0.510){87}{\rule{0.120pt}{0.509pt}}
\multiput(850.17,184.00)(-45.000,44.944){2}{\rule{0.400pt}{0.254pt}}
\multiput(804.92,230.00)(-0.498,0.510){87}{\rule{0.120pt}{0.509pt}}
\multiput(805.17,230.00)(-45.000,44.944){2}{\rule{0.400pt}{0.254pt}}
\multiput(758.45,276.58)(-0.643,0.498){67}{\rule{0.614pt}{0.120pt}}
\multiput(759.73,275.17)(-43.725,35.000){2}{\rule{0.307pt}{0.400pt}}
\multiput(713.39,311.58)(-0.662,0.498){65}{\rule{0.629pt}{0.120pt}}
\multiput(714.69,310.17)(-43.694,34.000){2}{\rule{0.315pt}{0.400pt}}
\multiput(668.45,345.58)(-0.643,0.498){67}{\rule{0.614pt}{0.120pt}}
\multiput(669.73,344.17)(-43.725,35.000){2}{\rule{0.307pt}{0.400pt}}
\multiput(622.34,380.58)(-0.984,0.496){43}{\rule{0.883pt}{0.120pt}}
\multiput(624.17,379.17)(-43.168,23.000){2}{\rule{0.441pt}{0.400pt}}
\multiput(577.34,403.58)(-0.984,0.496){43}{\rule{0.883pt}{0.120pt}}
\multiput(579.17,402.17)(-43.168,23.000){2}{\rule{0.441pt}{0.400pt}}
\multiput(531.32,426.58)(-1.293,0.498){67}{\rule{1.129pt}{0.120pt}}
\multiput(533.66,425.17)(-87.658,35.000){2}{\rule{0.564pt}{0.400pt}}
\multiput(439.09,461.58)(-1.978,0.496){43}{\rule{1.665pt}{0.120pt}}
\multiput(442.54,460.17)(-86.544,23.000){2}{\rule{0.833pt}{0.400pt}}
\multiput(343.13,484.58)(-3.857,0.492){21}{\rule{3.100pt}{0.119pt}}
\multiput(349.57,483.17)(-83.566,12.000){2}{\rule{1.550pt}{0.400pt}}
\put(176.0,496.0){\rule[-0.200pt]{21.681pt}{0.400pt}}
\put(689,23){\makebox(0,0){$\tau_1$}}
\put(980,23){\makebox(0,0){$\tau_0$}}
\put(1340,23){\makebox(0,0){$\tau$}}
\put(154,288){\makebox(0,0)[r]{$2a_1$}}
\put(154,392){\makebox(0,0)[r]{$3a_1$}}
\put(154,750){\makebox(0,0)[r]{$a$}}
\end{picture}
\begin{center}
Fig.1
\end{center}

\begin{thebibliography}{99}
\bibitem{GS} S.Giddings, A.Strominger.{\it Nucl. Phys.} {\bf B306} (1988) 890.
\bibitem{C} S.Coleman. {\it Nucl. Phys.} {\bf B307} (1988) 867.
\bibitem{GS2}
 S.Giddings, A.Strominger  {\it Nucl. Phys.} {\bf B307} (1988) 854.
\bibitem{LRT} G.V.Lavrelashvili, V.A.Rubakov, P.G.Tinyakov.
{\it Mod. Phys. Lett.} {\bf A3} (1988) 1231.
\bibitem{RT1} V.A.Rubakov, P.G.Tinyakov.
 {\it Pis'ma ZhETF (JETP Lett.)} {\bf 48} (1988) 297.
\bibitem{RT2} V.A.Rubakov, P.G.Tinyakov.
 {\it Phys. Lett.} {\bf B214} (1988) 334.
\bibitem{H} S.Hawking. In:{\it General Relativity, an Einstein centenary
survey}, Eds. S.W.Hawking, W.Israel. Cambridge, Cambridge University
Press, 1979.
\bibitem{ZJ} J.Zinn-Justin. {\it Phys. Rep.} {\bf 70} (1981) 109.
\bibitem{M} V.P.Maslov. {\it 
The Complex WKB Method for Nonlinear Equations.}
Moscow, Nauka, 1977.
\bibitem{L} K.Lee. {\it Phys. Rev. Lett.} {\bf 61} (1988) 263.
\bibitem{CL} S.Coleman, K.Lee. {\it Nucl. Phys.} {\bf B329} (1990) 387.
\end{thebibliography}
\end{document}